\pdfoutput=1
\newif\ifFull
\Fulltrue
\ifFull
\documentclass[11pt]{article}
\else
\documentclass{llncs}
\fi

\usepackage{graphicx}
\usepackage{amsmath}
\ifFull
\usepackage{amsthm}
\usepackage{fullpage}
\usepackage{hyperref}
\fi
\usepackage{amssymb}
\usepackage{times}
\usepackage{url}
\ifFull
\fi

\ifFull\else
\usepackage{cramped}
\renewcommand{\subsection}[1]{\paragraph{\textbf{#1}.}}
\fi

\ifFull
\newtheorem{theorem}{Theorem}[section]

\fi

\ifFull
\let\oldendproof\endproof
\def\endproof{\qed\oldendproof}
\fi

\newcommand{\abs}[1]{{\lvert#1 \rvert}}

\title{Extended $h$-Index Parameterized Data Structures for Computing Dynamic Subgraph Statistics}

\ifFull
\author{
David Eppstein \\
Dept. of Computer Science \\
\url{http://www.ics.uci.edu/~eppstein/}
\and
Michael T. Goodrich \\
Dept. of Computer Science \\
Univ. of California, Irvine \\
\url{http://www.ics.uci.edu/~goodrich/}
\and 
Darren Strash \\
Dept. of Computer Science \\
Univ. of California, Irvine \\
\url{http://www.ics.uci.edu/~dstrash/}
\and
Lowell Trott \\
Dept. of Computer Science \\
\url{http://www.ics.uci.edu/~ltrott/}
}
\date{}
\else
\author{David Eppstein \and Michael T. Goodrich \and Darren Strash \and Lowell Trott}
\institute{Computer Science Department, University of California, Irvine, USA.}
\fi

\begin{document}

\maketitle

\pagestyle{plain}

\begin{abstract} We present techniques for maintaining 
subgraph frequencies in a dynamic graph,
using data structures that are parameterized in terms of $h$, the 
\emph{$h$-index} of the graph.
Our methods extend previous results of Eppstein and Spiro for
maintaining statistics for undirected subgraphs of size three to
directed subgraphs and to subgraphs of size four.
For the directed case, we provide a data 
structure to maintain counts for all 3-vertex induced subgraphs 
in $O(h)$ amortized time per update.
For the undirected case, we maintain the counts of size-four subgraphs
in $O(h^2)$ amortized time per update.
These extensions enable a number of new applications in
Bioinformatics and Social Networking research.
\end{abstract}

\section{Introduction}
\ifFull
Deriving inspiration from work done on
fixed-parameter tractable algorithms for NP-hard problems (e.g.,
see~\cite{cllor-fpadf-08,dffht-fparp-05,df-fptcb-95,gghnw-cbfp-06,nr-oefpa-03}),
\else
Deriving inspiration from work done on
fixed-parameter tractable algorithms for NP-hard problems (e.g.,
see~\cite{dffht-fparp-05,df-fptcb-95,nr-oefpa-03}),
\fi
the area of
\emph{parameterized algorithm design} involves defining numerical
parameters for input instances, other than just the input size,
and designing data structures and algorithms whose performance
can be characterized in terms of those parameters.
The goal, of course, is to find useful parameters and
then design data structures and algorithms that are efficient for
typical values of those parameters 
(e.g., see~\cite{eg-snprn-08,es-hgadss-09}).
In this paper, we are interested in extending previous applications
of this approach in the context of dynamic subgraph statistics---where 
one maintains the counts of all (induced and non-induced) subgraphs 
of certain types---from undirected size-three
subgraphs~\cite{es-hgadss-09} to applications involving directed size-three
subgraphs and undirected subgraphs of size four.

Upon cursory examination this contribution may seem incremental, but these 
extensions allow for the possibility of significant computational improvement in several important applications.
For instance, in bioinformatics, statistics involving the frequencies of 
certain small subgraphs, called \emph{graphlets}, have been applied
to protein-protein interaction networks~\cite{tp-ubnfv-08,pcj-eegfd-06} 
and cellular networks~\cite{p-bncug-07}.
In these applications, the frequency statistics for the
subgraphs of interest
have direct bearing on biological network structure and function.
In particular, in these graphlets applications,
the undirected subgraphs of interest include 
one size-two subgraph (the 1-path), two size-three subgraphs
(the 3-cycle and 2-path), and six size-four subgraphs
(the 3-star, 3-path, triangle-plus-edge, 4-cycle, $K_4$ minus an
edge, and $K_4$), which we respectively illustrate later in 
Fig.~\ref{fig-quadrangles}
as 
$Q_4$, 
$Q_6$, 
$Q_7$, 
$Q_8$, 
$Q_9$, 
and $Q_{10}$.

In addition,
maintaining subgraph counts in a dynamic graph is of crucial
importance to statisticians and social-networking researchers using the 
\emph{exponential random graph model} 
\ifFull
(ERGM)~\cite{Fra-SN-91,RobMor-SN-07,Sni-JoSS-02,WasPat-Psy-96}
\else
(ERGM)~\cite{Fra-SN-91,Sni-JoSS-02,WasPat-Psy-96}
\fi
to generate random graphs.
ERGMs can be tailored to
generate random graphs that possess
specific properties, which makes ERGMs an ideal tool for Social
Networking research~\cite{WasPat-Psy-96,Sni-JoSS-02}.
This tailoring is accomplished by
a Markov Chain Monte Carlo (MCMC) method~\cite{Sni-JoSS-02},
which generates random 
graphs via a sequence of incremental changes.
These incremental changes are accepted or rejected based on
the values of subgraph statistics, which must
be computed exactly for each incremental change in order to
facilitate acceptance or rejection.
Thus, there is a need for dynamic graph statistics in ERGM
applications.

Typical graph attributes of interest in ERGM applications include
the frequencies of undirected stars and triangles, 
which are used in the triad model~\cite{FraStr-JASA-86}
to study friends-of-friends relationships, as well as
other more-complex subgraphs~\cite{SniPatRob-SM-06},
including undirected 4-cycles and two-triangles ($K_4$ minus an edge), 
and directed transitive triangles, which we illustrate as graph $T_9$
in Fig.~\ref{fig-triangles}.
Therefore, there is
a salient need for algorithms to maintain subgraph statistics
in a dynamic graph involving directed subgraphs of size three and
undirected subgraphs of size four.

Interestingly, extending the previous approach, of 
Eppstein and Spiro~\cite{es-hgadss-09},
for maintaining undirected size-three subgraphs to these new contexts involves
overcoming some algorithmic challenges.
The previous approach uses a parameterized algorithm
design framework for counting three-vertex induced 
subgraphs in a dynamic undirected graph. Their data structure
has running time $O(h)$ amortized time per graph
update (assuming constant-time hash table lookups), 
where $h$ is the largest integer such that there exists
$h$ vertices of degree at least $h$, which is a parameter 
known as the \emph{$h$-index} of the graph.
This parameter was introduced by Hirsch~\cite{Hir-PNAS-05} as a combined
way of measuring productivity and
impact in the academic achievements of researchers. 
\ifFull
In spite of its
drawbacks for this purpose~\cite{AdlEwiTay-JCQAR-08}, it is a useful
parameter for dynamic graph algorithms, as demonstrated 
by Eppstein and Spiro.
\fi
As we will show,
extending the approach of Eppstein and Spiro to directed
subgraphs of size three and undirected subgraphs of size four involves
more than doubling the 
complexity of the algebraic expressions and supporting data structures needed.
Ensuring the directed size-three procedure maintains the complexity bounds of 
previous work required extensive understanding of dynamic graph composition.  Developing 
the approach for size-four subgraphs that would allow only the addition of a single 
factor of $h$ required innovative work with the structure of stored graph elements.

\subsection{Other Related Work}
Although subgraph isomorphism is known to be NP-complete,
it is solvable in polynomial time for small subgraphs.
For example, all triangles and four-cycles can be found 
in an $n$-vertex graph with $m$ edges 
in $O(m^{3/2})$ time~\cite{ItaRod-SJC-78,ChiNis-SICOMP-85}. 
All cycles up to length seven can be counted (but not listed) in 
$O(n^\omega)$ time~\cite{AloYusZwi-Algo-97},
where $\omega\approx 2.376$ is the exponent for the asymptotically 
fastest known matrix multiplication algorithm~\cite{CopWin-JSC-90}.
\ifFull
In addition,
fast matrix multiplication has also been used for other problems of finding
and counting small cliques in graphs and
hypergraphs~\cite{EisGra-TCS-04,KloKraMue-IPL-00,NesPol-CMUC-85,VasWil-STOC-09,Yus-IPL-06}.
\fi
Also, in planar graphs, 
the number of copies of any fixed subgraph may be found in linear
time~\cite{Epp-JGAA-99,Epp-Algo-00}.
These previous approaches run too slowly for the iterative nature of
ERGM Markov Chain Monte Carlo simulations, however.

\subsection{Our Results}
In this paper,
we present an extension of the $h$-index parameterized data structure
design from statistics for undirected subgraphs of size three to
directed subgraphs of size three and undirected subgraphs of size
four.
We show that in a dynamic directed graph one can maintain the
counts of all directed three-vertex subgraphs in $O(h)$ 
amortized time per update, and in a dynamic undirected graph one
can maintain the four-vertex subgraph counts in $O(h^2)$ 
amortized time per update, assuming constant-time hash-table lookups
(or worst-case amortized times that are a logarithmic factor larger).
These results therefore provide techniques for application domains,
in Bioinformatics and Social Networking,
that can take advantage of these extended types of statistics.
In addition, our data structures are based a number of novel insights
into the combinatorial structure
of these different types of subgraphs.

\section{Preliminaries}
As mentioned above, we define the \emph{h-index}
of a graph to be the largest $h$ such that the graph contains $h$ vertices
of degree at least $h$. We define the $h$-partition of a graph to be
the sets $(H, V \setminus H)$, where $H$ is the set of vertices that 
form the $h$-index.

\subsection{The H-Index}

\begin{figure}[hb!]
\vspace{-24pt}
\begin{center}
\includegraphics[scale=0.95]{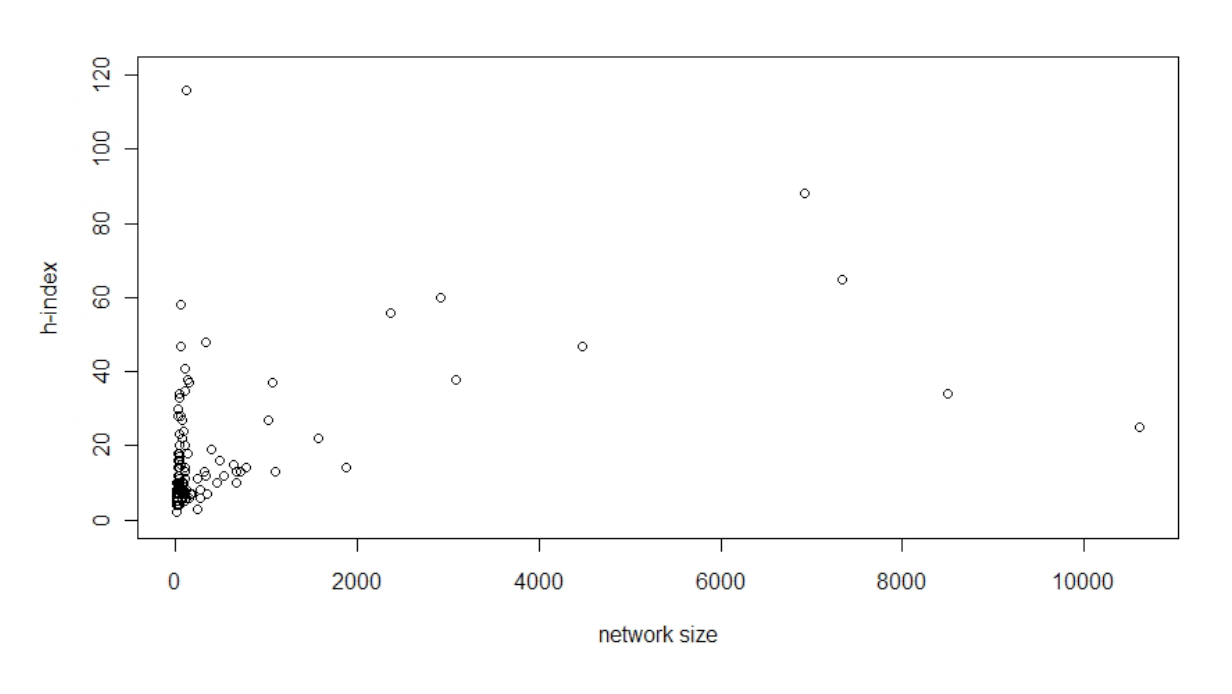}
\end{center}
\vspace{-24pt}
\caption{Scatter plot of h-index and network size from Eppstein and Spiro~\cite{es-hgadssarx-09}}
\label{fig-h-index}
\vspace{-18pt}
\end{figure}

\begin{figure}[hb!]
\vspace{-24pt}
\begin{center}
\includegraphics[scale=0.95]{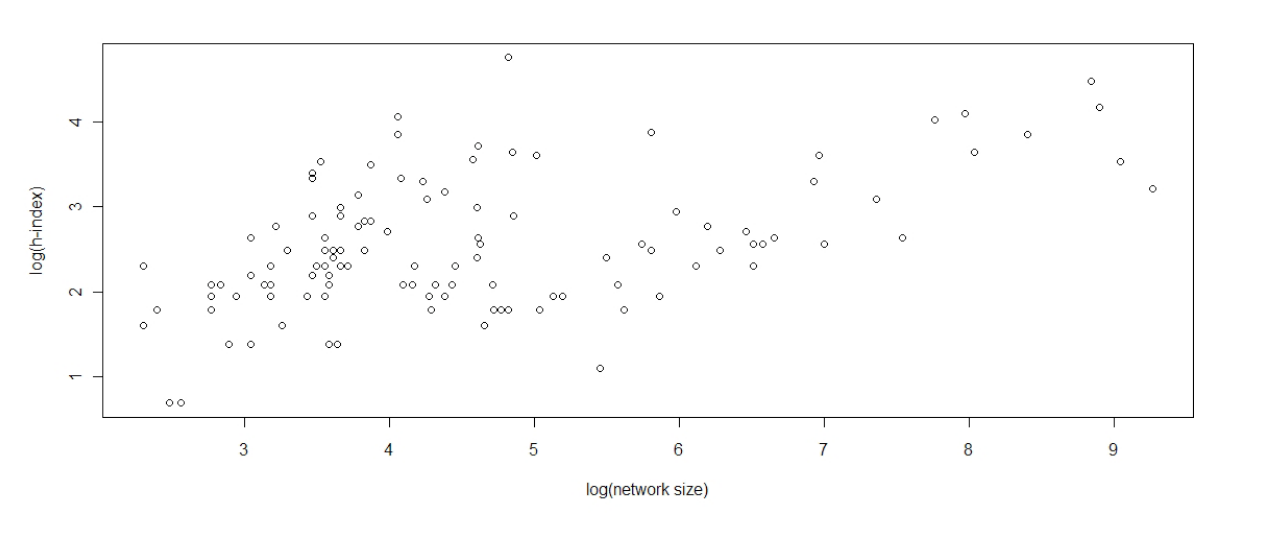}
\end{center}
\vspace{-24pt}
\caption{Scatter plot of h-index and network size, on log-log scale from Eppstein and Spiro~\cite{es-hgadssarx-09}}
\label{fig-h-loglog}
\vspace{-18pt}
\end{figure}

It is easy to see that the $h$-index of a graph with $m$ edges 
is $O(\sqrt{m})$; hence it is $O(\sqrt{n})$ for sparse graphs with a
linear number of edges, where $n$ is the number of vertices.
Moreover,
this bound is optimal in the worst-case, e.g., for a graph consisting
of $\sqrt{n}$ stars of size $\sqrt{n}$ each.
As can be seen in Fig.~\ref{fig-h-index} Eppstein and Spiro~\cite{es-hgadss-09} 
show experimentally that real-world social networks often have
$h$-indices much lower than the indicated worst-case bound.
These indices, perhaps more easily viewed in log-log scale in Fig~\ref{fig-h-loglog}, were calculated on
networks with a range of ten to just over ten-thousand nodes. The $h$-index of these networks were consistently below forty with only a few exceptions, 
none greater than slightly above one-hundred.
Moreover, many large real-world networks possess
\emph{power laws}, so that their number of vertices 
with degree $d$ is proportional to $nd^{-\lambda}$, for 
some constant $\lambda>1$.
Such networks are said to be 
\ifFull
\emph{scale-free}~\cite{AlbJeoBar-Nat-99,LilEdlAma-Nat-01,New-SIAM-03,Pri-Sci-65}, and it is often the case that
\else
\emph{scale-free}~\cite{AlbJeoBar-Nat-99,New-SIAM-03,Pri-Sci-65}, and it is often the case that
\fi
the parameter $\lambda$ is between $2$ and $3$ in real-world networks.
Note that the $h$-index of a scale-free graph is
 $h=\Theta(n^{1/(1+\lambda)})$, since it must satisfy the equation
 $h=nh^{-\lambda}$.
 Thus, for instances of scale-free graphs with $\lambda$ between $2$
 and $3$, an algorithmic performance of $O(h)$ is much better than
  the worst-case $O(\sqrt n)$ bound for graphs without power-law degree 
  distributions. For example, an $O(h)$ time bound for a scale-free
  graph with
  $\lambda= 2$ would give a bound of $O(n^{1/3})$ while for $\lambda=3$
  it would give an $O(n^{1/4})$ bound. 
  Likewise, an algorithmic performance of $O(h^2)$ is much better
  than a worst-case performance of $O(n)$ for these instances, for
  $\lambda= 2$ would give a bound of $O(n^{2/3})$ while for $\lambda=3$
  it would give an $O(n^{1/2})$ bound. 
Thus, by taking a parametric algorithm design approach, we can,
in these cases, achieve running times
better than worst-case bounds characterized strictly
in terms of the input size, $n$.

\subsection{Maintaining Undirected Size-3 Subgraph Statistics}
As mentioned above,
Eppstein and Spiro~\cite{es-hgadss-09} 
develop an algorithm for maintaining the $h$-index and 
the $h$-partition of a graph among edge insertions, edge deletions,
and insertions/deletions of isolated vertices in constant time plus a constant
number of dictionary operations per update. Observing that the $h$-index 
doubles after $\Omega(h^2)$ updates, Eppstein and Spiro further show a 
partitioning scheme requiring amortized $O(1/h)$ partition changes per 
graph update. This partitions the graph into sets of \emph{low}- and
\emph{high}-degree vertices, which we summarize in Theorem~\ref{thm-partition}. 

\begin{theorem}[\cite{es-hgadss-09}]
\label{thm-partition}
For a dynamic graph $G=(V,E)$, we can maintain a partition $(H, V\setminus H)$ 
such that for $v\in H$, $\mathrm{degree}(v) = \Omega(h)$ and $\abs{H} = O(h)$;
and for $u \in V\setminus H$, $\mathrm{degree}(u) = O(h)$ in constant time per
update, with amortized $O(1/h)$ changes to the partition per update.
\end{theorem}

Using this partitioning scheme, one can develop a triangle-counting 
algorithm as follows. For each pair of vertices $i$ and $j$, store 
the number of length-two paths $P[i,j]$ that have an intermediate 
low-degree vertex. Whenever an edge $(u,v)$ is added to the graph, 
increase the number of triangles by $P[u,v]$, and update the number of
length-two paths containing $(u,v)$ in $O(h)$ time. One can
then iterate over all the high-degree vertices, adding to a 
triangle count when a 
high-degree vertex is adjacent to both $u$ and $v$. Since there are 
$O(h)$ high-degree vertices, this method takes $O(h)$ time. These same steps 
can be done in reverse for an edge removal.

Whenever the partition changes, one must update $P[\cdot,\cdot]$ values
to reflect vertices moving from high to low, or low to high, which 
requires $O(h^2)$ time. Since there are amortized $O(1/h)$ partition 
changes per graph update, this updating takes $O(h)$ amortized time 
per update. The randomization comes from the choice of dictionary 
scheme used. The data structure as described requires $O(mh)$ space, 
which is sufficient to store the length-two paths with an intermediate 
low-degree vertex. 

Finally, to maintain counts of all induced undirected 
subgraphs on three vertices,
it suffices to solve a simple four-by-four 
system of linear equations relating induced
subgraphs and non-induced subgraphs. This allows one to keep counts 
of the induced subgraphs of every type with a constant amount of 
work in addition to counting triangles.
Extending this to directed subgraphs of size three and undirected
subgraphs of size four requires that we come up with a much larger
system of equations, which characterize the combinatorial
relationships between such types of subgraphs.

\section{Directed Three-Vertex Induced Subgraphs}

\begin{figure}[hb!]
\vspace{-24pt}
\begin{center}
\includegraphics[scale=0.45]{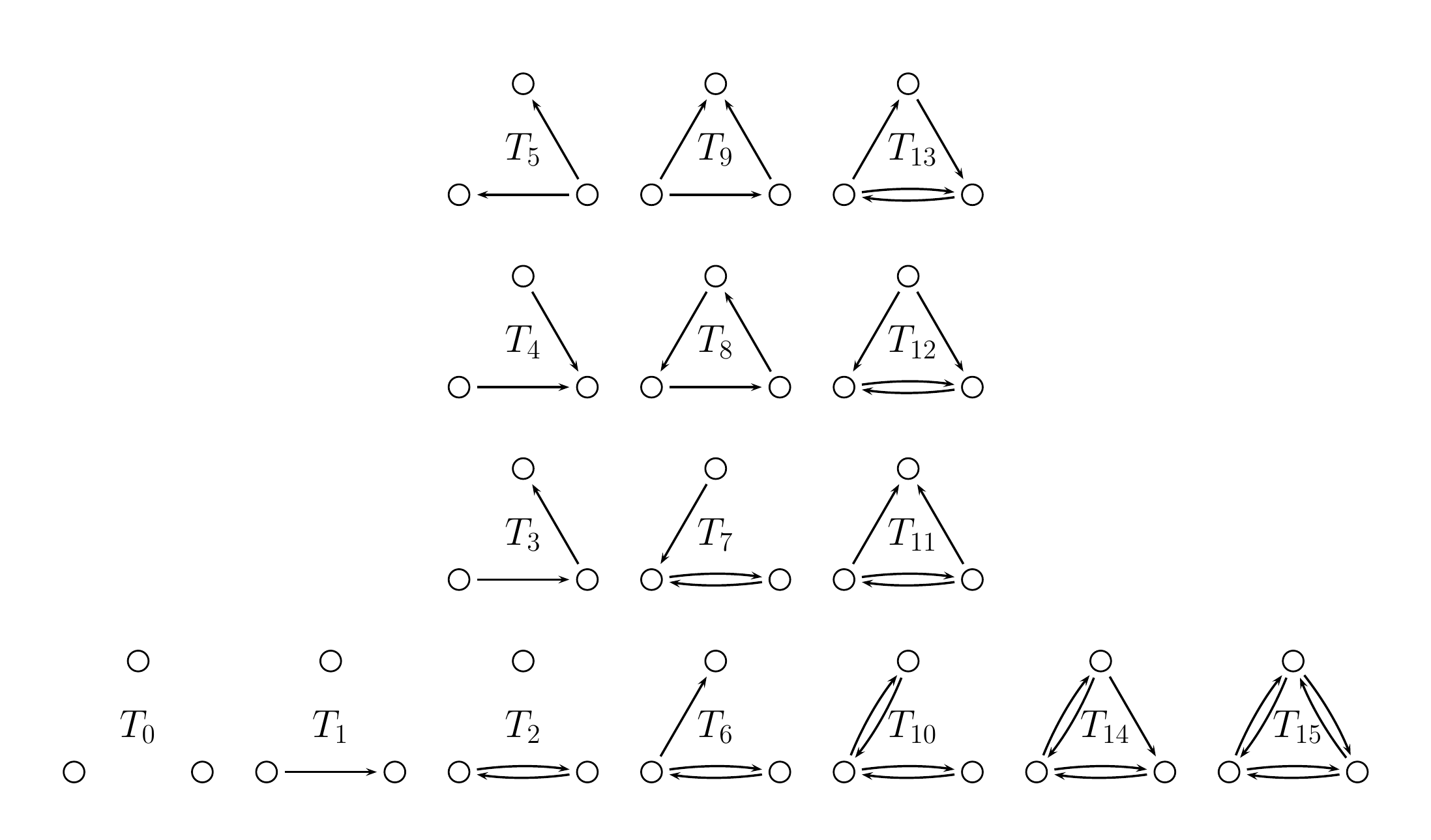}
\end{center}
\vspace{-24pt}
\caption{The 16 possible directed graphs on three vertices, excluding 
isomorphisms, organized in left-to-right order by number of edges in the graph. We
label these graphs $T_0$ to $T_{15}$.}
\label{fig-triangles}
\vspace{-18pt}
\end{figure}

Using the partitioning scheme detailed in Theorem~\ref{thm-partition}, we
can maintain counts for the all possible induced subgraphs on three vertices (see 
Fig.~\ref{fig-triangles}) in $O(h)$ amortized time per update for a dynamic
directed graph. We begin by maintaining counts for induced subgraphs that are a
directed triangle, we then show how to maintain counts of all induced
subgraphs on three vertices.

\subsection{Counting Directed Triangles}

\begin{figure}[hb!]
\vspace{-24pt}
\begin{center}
\includegraphics[scale=0.45]{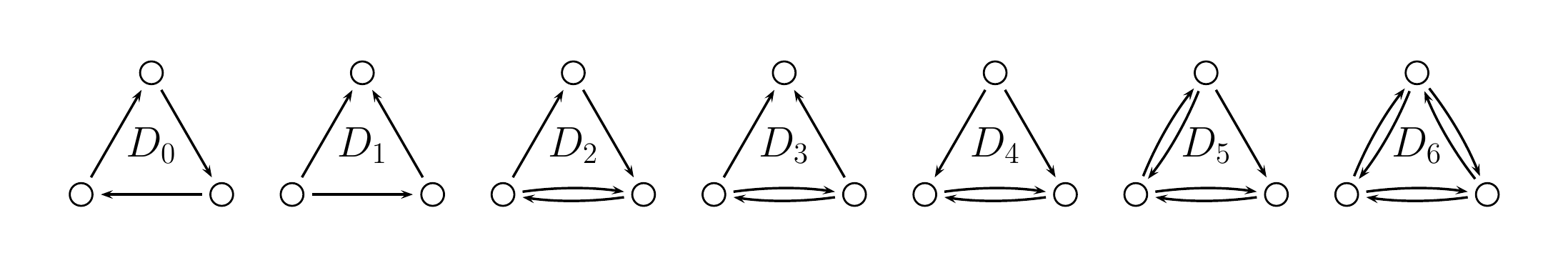}
\end{center}
\vspace{-24pt}
\caption{The 7 directed triangles, labeled $D_0$ to $D_6$.}
\label{fig-ditriangles}
\vspace{-12pt}
\end{figure}

Let a \emph{directed triangle} be a three-vertex directed graph 
with at least one directed edge between each pair of vertices. There
are seven possible directed triangles, labeled $D_0$ to $D_6$ in 
Fig.~\ref{fig-ditriangles}. We let $d_k$ denote the count of induced directed 
triangles of type $D_k$ in the dynamic graph. We now show how to maintain  
each count $d_i$ by extending Eppstein and Spiro's technique.

\ifFull
When an edge $(u,v)$ is added or removed from the graph, we would like 
to quickly compute the number of directed triangles containing $(u,v)$,
in order to update the current counts. The third vertex of this
directed triangle can either be low- or high-degree. We handle these
cases separately.\fi

\begin{figure}[t!]
\vspace{-24pt}
\begin{center}
\includegraphics[scale=0.45]{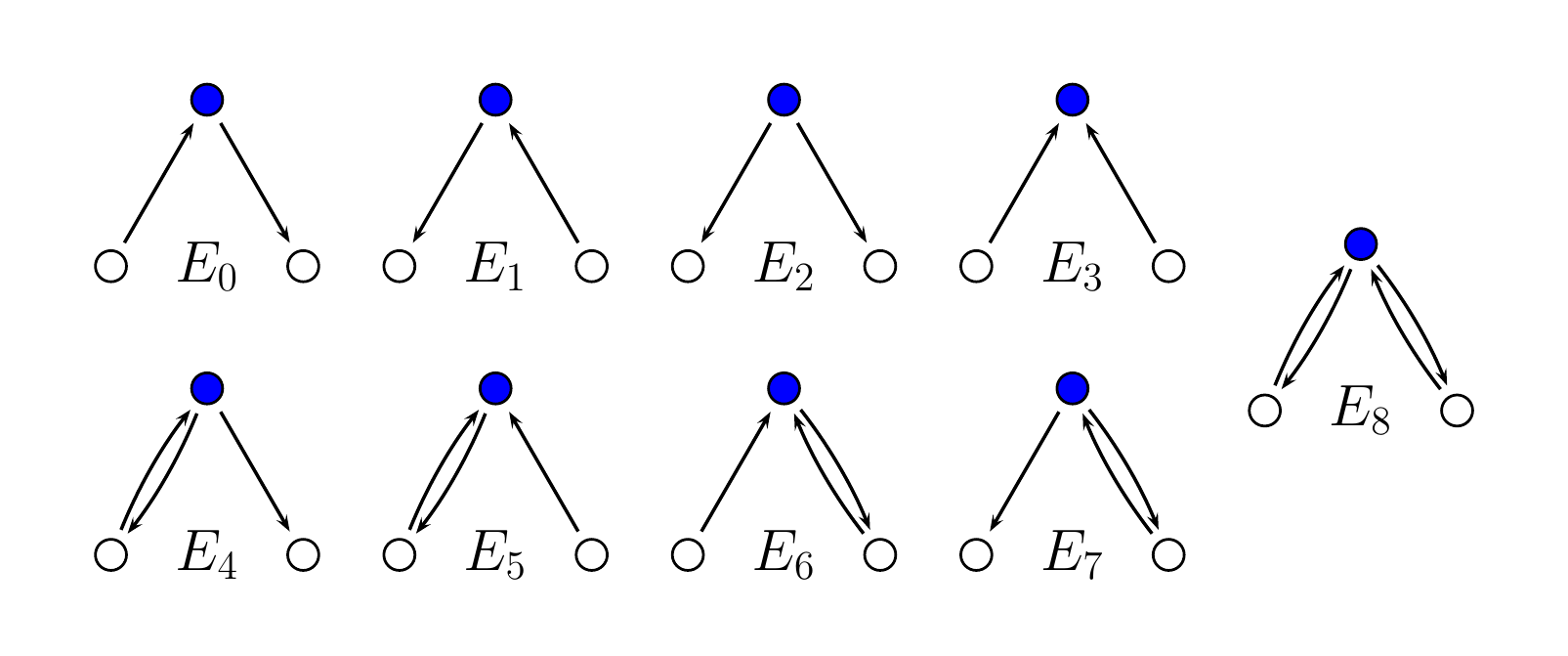}
\end{center}
\vspace{-24pt}
\caption{The nine elbows with a fixed joint.}
\label{fig-elbows}
\vspace{-12pt}
\end{figure}

For a pair of vertices $i$ and $j$, we define a \emph{joint} to 
be a third vertex $l$ that is adjacent to both $i$ and $j$. Vertices $i$, 
$l$ and $j$ are said to form an \emph{elbow}. Fixing a vertex to be a joint, 
there are nine unique elbows which we label $E_0$ to 
$E_8$(see Fig.~\ref{fig-elbows}). We store a dictionary mapping pairs of 
vertices $i$ and $j$ to the number of elbows of type $E_k$ formed by $i$ and $j$ and a 
low-degree joint, denoted $e_k[i,j]$. 

We now discuss how the directed triangle counts change when adding an edge
$(u,v)$. We do not discuss edge removal since its effects are symmetric 
to edge insertion.

For directed triangles with a third low-degree vertex, we update our counts 
using the dictionary of elbow counts. If edge $(v,u)$ is 
not in the graph, directed triangle counts increase as follows.

\vspace*{-8pt}
{\small
\begin{align*}
d_0 &= d_0 + e_1[u,v]\\
d_1 &= d_1 + e_0[u,v] + e_2[u,v] + e_3[u,v]\\
d_2 &= d_2 + e_5[u,v] + e_7[u,v]\\
d_3 &= d_3 + e_4[u,v]\\
d_4 &= d_4 + e_6[u,v]\\
d_5 &= d_5 + e_8[u,v]
\end{align*}
}

\vspace*{-16pt}
If edge $(v,u)$ is present in the graph, adding $(u,v)$ destroys
some directed triangles containing $(v,u)$. Therefore, the directed 
triangle counts change as follows.

\vspace*{-8pt}
{\small
\begin{align*}
d_0 &= d_0 - e_1[v,u]\\
d_1 &= d_1 - (e_0[v,u] + e_2[v,u] + e_3[v,u])\\
d_2 &= d_2 + (e_0[u,v] + e_1[u,v]) - (e_5[v,u] + e_7[v,u])\\
d_3 &= d_3 + e_3[u,v] - e_4[v,u]\\
d_4 &= d_4 + e_2[u,v] - e_6[v,u]\\
d_5 &= d_5 + (e_4[u,v] + e_5[u,v] + e_6[u,v] + e_7[u,v]) - e_8[v,u]\\
d_6 &= d_6 + e_8[u,v]
\end{align*}
}
\vspace*{-16pt}

To complete the directed triangle counting step, we iterate over the 
$O(h)$ high-degree vertices to account for directed triangles formed with $u$ 
and $v$ and a high-degree vertex, taking $O(h)$ time.

If either $u$ or $v$ is a low-degree vertex, we must also update the elbow 
counts involving the added edge $(u,v)$. We consider, without loss of generality,
the updates when $u$ is considered the low-degree elbow joint. For ease of notation,
we categorize the different relationships between adjacent vertices as follows:

\vspace*{-6pt}
{\small
\begin{align*}
\mathrm{inneighbor}(u) &= \{w\in V: (w,u)\in E \wedge (u,w) \not\in E\} \\
\mathrm{outneighbor}(u) &= \{w\in V: (u,w)\in E \wedge (w,u) \not\in E\} \\
\mathrm{neighbor}(u) &= \{w\in V: (u,w)\in E \wedge (w,u) \in E\}.
\end{align*} 
}

\vspace*{-6pt}
We summarize the elbow count updates in Table~\ref{tab-elbowupdate}.

\begin{table}[!b]
\vspace*{-12pt}
\caption{Summary of updating elbow counts when $u$ is considered a low-degree joint.}
\label{tab-elbowupdate}
\begin{center}
\begin{tabular}{l|c|c}
& $(v,u) \not\in E$ & $(v,u) \in E$ \\ \hline
$w \in \mathrm{inneighbor}(u)\setminus \{v\}$ &
$\begin{aligned}
e_0[w,v] &= e_0[w,v] + 1\\
e_1[v,w] &= e_1[v,w] + 1
\end{aligned}$ & 

$\begin{aligned}
e_6[w,v] &= e_6[w,v] + 1\\
e_5[v,w] &= e_5[v,w] + 1
\end{aligned}$ \\ \hline

$w \in \mathrm{outneighbor}(u)\setminus \{v\}$ &
$\begin{aligned}
e_0[v,w] &= e_0[v,w] + 1\\
e_1[w,v] &= e_1[w,v] + 1
\end{aligned}$ & 

$\begin{aligned}
e_4[v,w] &= e_4[v,w] + 1\\
e_7[w,v] &= e_7[w,v] + 1
\end{aligned}$ \\ \hline

$w \in \mathrm{neighbor}(u)\setminus \{v\}$ &
$\begin{aligned}
e_4[w,v] &= e_4[w,v] + 1\\
e_7[v,w] &= e_7[v,w] + 1
\end{aligned}$ & 

$\begin{aligned}
e_8[w,v] &= e_8[w,v] + 1\\
e_8[v,w] &= e_8[v,w] + 1
\end{aligned}$ \\
\end{tabular}
\end{center}
\end{table}

Finally, when there is a partition change, we must update the elbow counts.
If node $w$ moves across the partition, then we consider all pairs of neighbors of 
$w$ and update their elbow counts appropriately. Since there are $O(h^2)$ pairs
of neighbors, and a constant number of elbows, this step takes $O(h^2)$ time. Since
$O(1/h)$ amortized partition changes occur with each graph update, this step requires
$O(h)$ amortized time.

\subsection{Subgraph Multiplicity}

Let the count for induced subgraph $T_i$ be called $t_i$.  Furthermore, for
a vertex $v$, let $i(v)=\abs{\mathrm{inneighbor}(v)}$, 
$o(v)=\abs{\mathrm{outneighbor}(v)}$ and $r(v) = \abs{\mathrm{neighbor}(v)}$.
We can represent the relationship between the number of induced and non-induced 
subgraphs using the matrix equation

\vspace*{-10pt}
{\small
\setcounter{MaxMatrixCols}{16}
\[
\begin{bmatrix}
1&1&1&1&1&1&1&1&1&1&1&1&1&1&1&1 \\
0&1&2&2&2&2&3&3&3&3&4&4&4&4&5&6 \\
0&0&1&0&0&0&1&1&0&0&2&1&1&1&2&3 \\
0&0&0&1&0&0&1&1&3&1&2&2&2&3&4&6 \\
0&0&0&0&1&0&0&1&0&1&1&1&2&1&2&3 \\
0&0&0&0&0&1&1&0&0&1&1&2&1&1&2&3 \\
0&0&0&0&0&0&1&0&0&0&2&2&0&1&3&6 \\
0&0&0&0&0&0&0&1&0&0&2&0&2&1&3&6 \\
0&0&0&0&0&0&0&0&1&0&0&0&0&1&1&2 \\
0&0&0&0&0&0&0&0&0&1&0&2&2&1&3&6 \\
0&0&0&0&0&0&0&0&0&0&1&0&0&0&1&3 \\
0&0&0&0&0&0&0&0&0&0&0&1&0&0&1&3 \\
0&0&0&0&0&0&0&0&0&0&0&0&1&0&1&3 \\
0&0&0&0&0&0&0&0&0&0&0&0&0&1&2&6 \\
0&0&0&0&0&0&0&0&0&0&0&0&0&0&1&6 \\
0&0&0&0&0&0&0&0&0&0&0&0&0&0&0&1
\end{bmatrix}
\begin{bmatrix}
t_0 \\ t_1 \\ t_2 \\ t_3 \\ t_4 \\ t_5 \\ t_6 \\ t_7 \\ t_8 \\ 
t_9 \\ t_{10} \\ t_{11} \\ t_{12} \\ t_{13} \\ t_{14} \\ t_{15}
\end{bmatrix}
=
\begin{bmatrix}
n_0 = \binom{n}{3} \\
n_1 = m(n-2) \\
n_2 = \frac{1}{2}(n-2)\sum_{v\in V}r(v) \\
n_3 = \sum_{(u,v)\in E}\sum_{(v,w)\in E, w\neq u}1 \\
n_4 = \sum_{v\in V}\binom{\mathrm{indegree}(v)}{2} \\
n_5 = \sum_{v\in V}\binom{\mathrm{outdegree}(v)}{2} \\
n_6 = \sum_{v\in V}(\binom{r(v)}{2} + \mathrm{o}(v)\cdot r(v)) \\
n_7 = \sum_{v\in V}(\binom{r(v)}{2} + \mathrm{i}(v)\cdot r(v)) \\
n_8 = d_0 + d_2 + d_5 + 2d_6 \\
n_9 = d_1 + d_2 + 2d_3 + 2d_4 + 3d_5 + 6d_6 \\
n_{10} = \sum_{v\in V}\binom{r(v)}{2} \\
n_{11} = d_3 + d_5 + 3d_6 \\
n_{12} = d_4 + d_5 + 3d_6 \\
n_{13} = d_2 + 2d_5 + 6d_6 \\
n_{14} = d_5 + 6d_6 \\
n_{15} = d_6
\end{bmatrix}.
\]
}

On the right hand side, each $n_i$ is the count of the number of
non-induced $T_i$ subgraphs in the dynamic graph. Each $n_i$ (excluding
directed triangle counts) is maintained in constant time per update by 
storing a constant amount of structural information at each node, 
such as indegree, outdegree, and reciprocity of neighbors. On the left 
hand side, position $i,j$ in the matrix counts how many non-induced 
subgraphs of type $T_i$ appear in $T_j$. We are counting non-induced 
subgraphs in two ways: (1) by counting the number of appearances within 
induced subgraphs and (2) by using the structure of the graph. Since 
the multiplicand is an upper (unit) triangular matrix, this matrix 
equation is easily solved, yielding the induced subgraph counts. Thus,
we can maintain the counts for three-vertex induced subgraphs in a directed
dynamic graph in $O(h)$ amortized time per update, and $O(mh)$ space, 
plus the additional overhead for the choice of dictionary.

\section{Four-Vertex Subgraphs}

\begin{figure}[!b]
\vspace{-20pt}
\begin{center}
\includegraphics[scale=0.45]{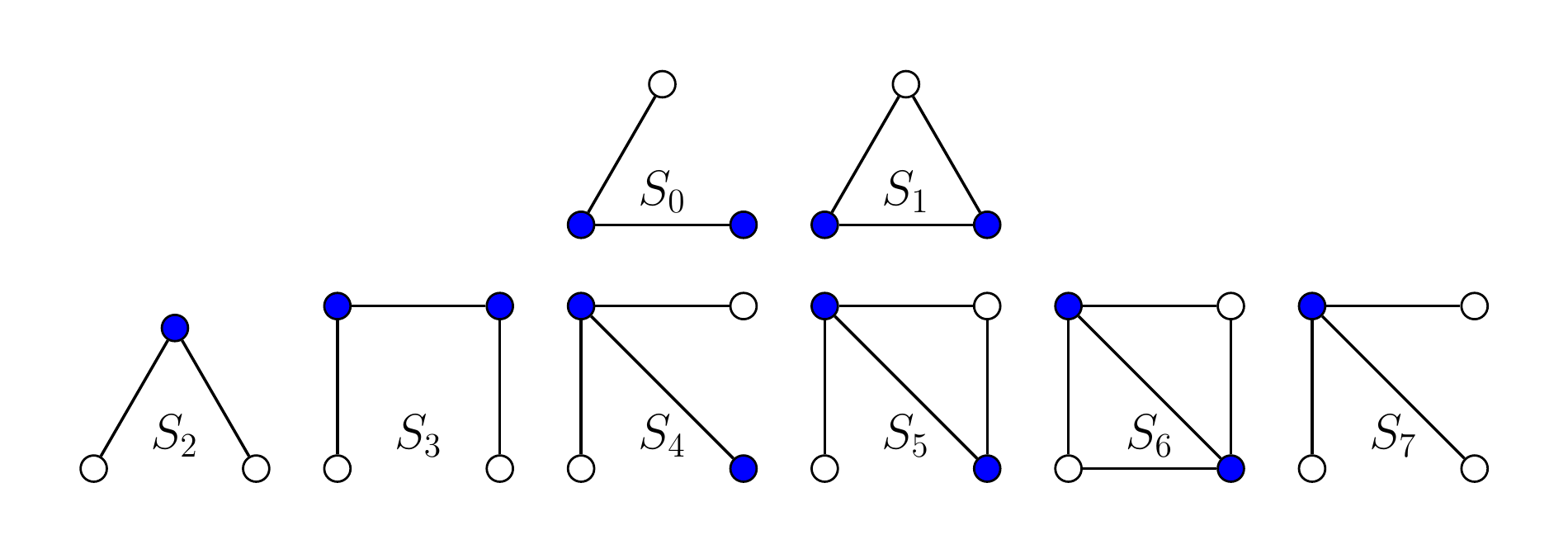}
\end{center}
\vspace{-24pt}
\caption{We store counts of these eight non-induced subgraphs to maintain counts of
four-vertex non-induced subgraphs $Q_3$ to $Q_{10}$. The counts are indexed by the
labels of the white vertices, and the blue vertices indicate a vertex 
has low-degree.}
\label{fig-quadstore}
\end{figure}

We begin by describing the data structure for our algorithm.  It will be necessary to maintain the counts of various subgraph structures.  The data structure in whole consists of the following information:

\begin{itemize}
\item Counts of the non-induced subgraph structures, $m_3$ through $m_{10}$.

\item A set E of the edges in the graph, indexed such that given a pair of endpoints there is a constant-time lookup to determine if they are linked by an edge.

\item A partition of the vertices of the graph into two sets $H$ and $V \setminus H$. 

\item A dictionary $P_1$ mapping each vertex $u$ to a pair $P_1[u] = (s_0[u]$, $s_1[u])$.  This pair contains the counts for the structures $S_0$ and $S_1$ that involve vertex $u$ ( see  Fig.~\ref{fig-quadstore}). That is, the count of the number of two-edge paths that begin at $u$ and pass through two vertices in $V \setminus H$ and the number of these paths that connect back to $u$ forming a triangle.  We only maintain nonzero values for these numbers in $P_1$; if there is no entry in $P_1[u]$ for the vertex $u$ then there exist no such path from $u$.

\item A dictionary $P_2$ mapping each pair of vertices $u$, $v$ to a tuple $P_2[u, v] = (s_2[u,v]$, $s_3[u,v]$, $s_4[u,v]$, $s_5[u,v]$, $s_6[u,v])$.  This tuple contains the counts for the structures $S_2$ through $S_6$ that involve vertices $u$ and $v$ ( see  Fig.~\ref{fig-quadstore}).  That is, the number of two-edge paths from u to v via a vertex of $V \setminus H$, the number of three-edge paths from u to v via two vertices of $V \setminus H$, the number of structures in which both $u$ and $v$ connect to the same vertex in $V \setminus H$ which connects to another vertex in $V \setminus H$, the number of structures similar to the last in which the final vertex in $V \setminus H$ shares an edge connection with $u$ or $v$, and the number of structures where between $u$ and $v$ there are two two-edge paths through vertices of $V \setminus H$ in which the two vertices in $V \setminus H$ share an edge connection.  Again, we only maintain nonzero values.

\item A dictionary $P_3$ mapping each triple of vertices $u$, $v$, $w$  to a number $P_2[u, v, w] = (s_7[u,v,w])$.  This value is the count for the structure $S_7$ that involves vertices $u$, $v$, and $w$ ( see  Fig.~\ref{fig-quadstore}). This is, the number of vertices in $V \setminus H$ that share edge connections with all three vertices.  As before, we only maintain nonzero values for these numbers.

\end{itemize}

\begin{figure}
\vspace{-24pt}
\begin{center}
\includegraphics[scale=0.45]{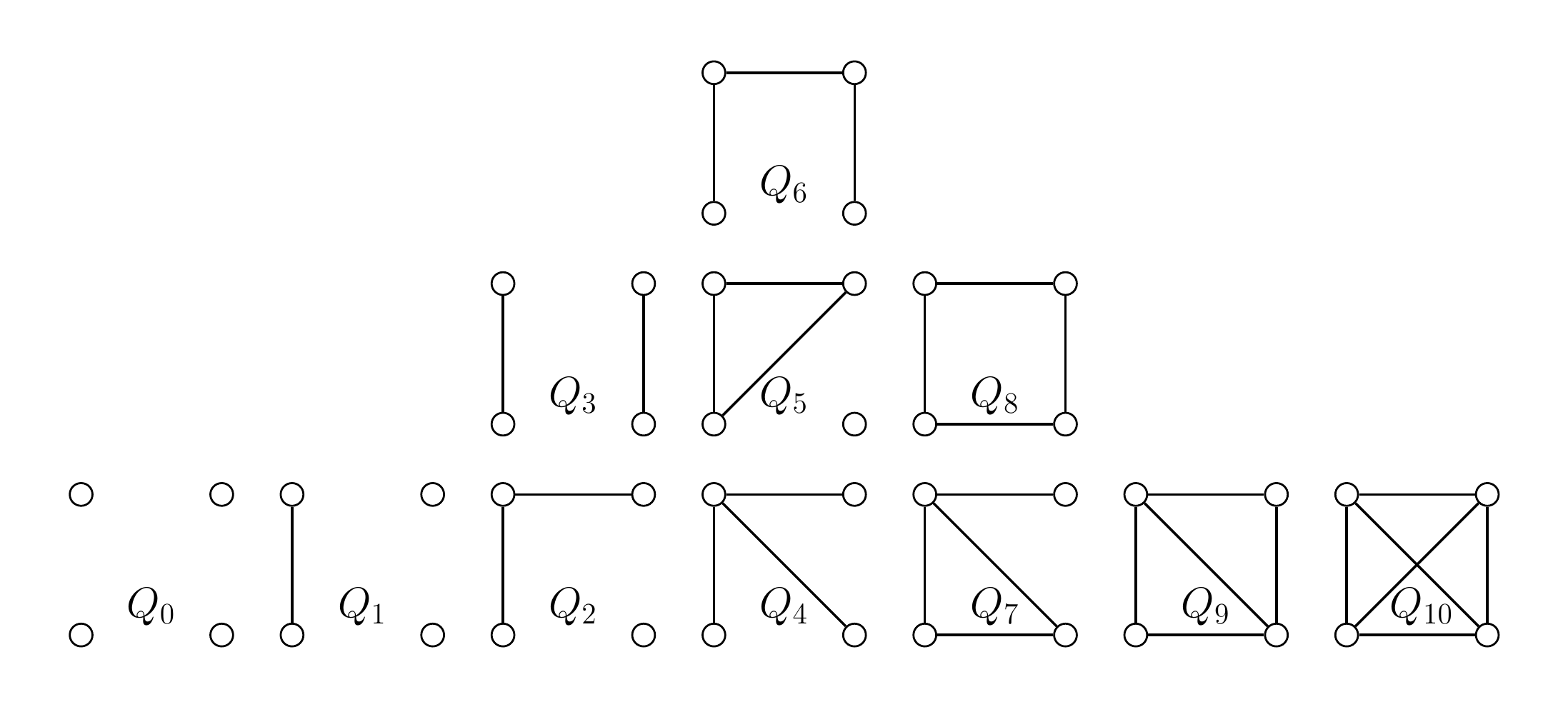}
\end{center}
\vspace{-24pt}
\caption{The 11 possible graphs on four vertices, excluding 
isomorphisms, organized in left-to-right order by number of edges in the graph.}
\label{fig-quadrangles}
\end{figure}

Upon insertion of an edge between vertices $v_1$ and $v_2$ we will need to update the dictionaries $P_1$, $P_2$, and $P_3$. If both $v_1$ and $v_2$ are in $H$, no update is necessary.

If $v_1$ and $v_2$ are both in $V \setminus H$ then we will need to update the counts $s_0$ through $s_6$.  First find which vertices in $H$ connect to $v_1$ or to $v_2$.  Increment $s_0$ for these vertices.  If both vertices in $V \setminus H$ connect to the same vertex in $H$ then increment $s_1$ for this vertex.  Increment $s_2$ for $v_1$ and the vertices that connect to $v_2$, and for $v_2$ and the vertices that connect to $v_1$.  Then increment $s_3$ based on pairs of neighbors of $v_1$ and $v_2$ and neighbors of neighbors in $V \setminus H$.  If either $v_1$ or $v_2$ connect to two vertices in $H$ increment $s_4$ for the vertices in $H$.  Considering $v_1$ to be the vertex with edge connections to two vertices in $H$, for each vertex in $H$ that connects to $v_2$ increment $s_5$.  For two vertices in $H$ such that $v_1$ and $v_2$ each connect to both, increment $s_6$ for the vertices in $H$.  

If $v_1$ and $v_2$ are such that one is in $V \setminus H$ and the other in $H$ we will proceed as follows.  Consider $v_1$ to be the vertex in $V \setminus H$. First, determine the number of vertices in $V \setminus H$ connected to $v_1$ and increase $s_0$ for $v_2$ by that amount.  Upon discovering these adjacent vertices in $V \setminus H$ test their connection to $v_2$.  For each of those connected, increment $s_1$ for $v_2$.  It is necessary to determine which vertices in $H$ share an edge with $v_1$.  After these connections have been determined increment the appropriate dictionary entries.  Form pairs with $v_2$ and the connected vertices in $H$ and update the $s_2$ counts.  Form triples with $v_2$ and two other connected vertices in $H$ and update the counts in $s_7$. The $s_5$ update comes from determining the triangles formed by the additional edge and using the degree of the vertices in $H$, and the count of the connected triangles, which can be calculated by searching for attached vertex pairs in $H$ and using $s_2$. In order to update the count for $s_6$ begin with location of vertex pairs as with the elbow update.  For each of the $H$ vertex pairs increase the stored value by the number of vertices in $V \setminus H$ that share an edge with $v_1$ and with both of the vertices in $H$, which can be retrieved from $s_2$.

Examining the time complexity we can see that in order to generate the dictionary updates the most complex operation involves examination of two sets of connected vertices consecutively that are $O(h)$ in size each.  This results in $O(h^2)$ operations to determine which updates are necessary.  Since it is possible to see from the structure of the stored items that no single edge insertion can result in more than $O(h^2)$ new structures, this will be the upper bound on dictionary updates, and make $O(h^2)$ the time complexity bound.

These maintained counts will have to be modified when the vertex partition is updated. If a vertex is moved from $H$ to $V \setminus H$ then it is necessary to count the connected structures it now forms.  This can be done by examining all edges formed by this vertex, and following the procedure for edge additions.  
When a vertex is moved into $H$ it is necessary to count the structures it had been forming as a vertex in $V \setminus H$ and decrement the appropriate counts.  This can be done similarly to the method for generating new structures.  In analysis of the partition updates we see that since we are working with a single vertex with $O(h)$ degree the complexity has an additional $O(h)$ factor to use the edge-based dictionary update scheme. This results in $O(h^3)$ time per update.  Since this partition update is done an average of $O(1/h)$ times per operation, the amortized time for updates, per change to the input graph, is $O(h^2)$.

\subsection{Subgraph Structure Counts}

The following section covers the update of the subgraph structure counts after an edge between vertices $v_1$ and $v_2$ has been inserted.  Let these vertices have degree count $d_1$ and $d_2$ respectively. Recall that $m_i$ refers to the count of the non-induced subgraph of the structure $Q_i$ (see Fig.~\ref{fig-quadrangles}).

The $m_3$ count will be increased by $(m-(d_1+d_2-2))$, where $m$ is the number of edges in the graph.  Since this structure consists of two edges that do not share vertices, the increase of the count comes from a selection of a second edge to be paired with the inserted edge.  The second term in the update value reflects the number of edges that connected to the inserted edge.

The $m_4$ count will be updated as follows.  Each of the two vertices can be the end of a claw structure.  From each end two edges in addition to the newly inserted edge must be selected.  Thus the value to update the count is $\binom{d_1-1}{2} + \binom{d_2-1}{2}$. 

The $m_5$ count is updated by calculating the number of additional triangles the edge addition would add, which can be done with the 
Eppstein-Spiro~\cite{es-hgadss-09}
method, and multiplying that by a factor of $(n-3)$ to reflect the selection of the additional vertex, where $n$ is the number of vertices in the graph.

The update for $m_6$ is done in parts based on which position in the structure the edge is forming.  The increase to the count for the new structures in which the additional edge is the center in the three-edge path is $((d_1 - 1)(d_2-1))$.  

This value will be increased by the count when the new edge is not the center of the structure.  The process to calculate the count increase will assume that $v_1$ connects to the rest of this structure.  The same process can be done without loss of generality with the assumption $v_2$ connects to the rest of the structure.  These values will then both be added to form the final part of the count update.  If $v_1$ is an element of $H$ then we will sum the results from the following subcases.  First we consider the case where the vertex adjacent to $v_1$ is in $H$.  The number of these paths of length two originating at $v_1$ can be counted by summing the degree of these vertices minus 1. We must also subtract one for each of the adjacent vertices in $H$ that are adjacent to $v_2$.  If $v_1$ is not an element of $H$, then it has h or less neighbors.  Sum over all neighbors the following value.  If the vertex does not have an edge connecting it to $v_2$ then the degree of the vertex; if it does the degree minus one.

The $m_7$ count is updated as follows.  An inserted edge can form the structure in three positions, so our final update will be the sum of those three counts.  For the first case let the inserted edge be the additional edge connected to the triangle.  For this case, we must do all of the following for both vertices and sum the result.  If the vertex is in $H$ retrieve $s_1$.  This gives us the connected triangles through vertices in $V \setminus H$.  Then determine which vertices in $H$ connect to the vertex.  Form the triangle counts with all vertices in $H$. Form those with one additional vertex in $H$ using $s_2$.  If the vertex is in $V \setminus H$, then determine its neighbors connections and form a connected triangle count.

In the second case the edge is in the triangle and shares a vertex with the additional edge.  The count can be determined in two parts.  First the triangles.  If either $v_1$ or $v_2$ are in $V \setminus H$ then the triangle count can be calculated.  If both are in $H$ then a lookup to $s_2$ will determine the number of triangles. The number of additional edges can then be calculated using the degrees of the vertices of the inserted edge, with care to not count the edges used to form the triangle.  The product of the triangle and additional edge will form the increase for this case.

The final case occurs when the inserted edge is part of the triangle, but does not share a vertex with the additional edge. If either $v_1$ or $v_2$ are in $V \setminus H$ then the triangle count can be calculated, and the degree of the vertices used to form these triangles can be used to calculate the count increase.  If both  $v_1$ or $v_2$ are in $H$ then there are three remaining subcases.  The count if all vertices are in $H$ can be determined.  If the vertex on the additional edge that is not in the triangle is in $H$, then using the three known vertices in $H$ and a lookup from $P_2$ can yield the counts.  If both remaining vertices are in $V \setminus H$ this is the structure stored in $s_4$, and counts can be retrieved.  Sum the counts for these subcases to calculate the total increase for this case.

The count for $m_8$ is increased upon edge update by a sum of the following.  The count of the length three path through vertices in $V \setminus H$ can be looked up in $s_3$.  There are two possible types of length three paths remaining.  In the first, both vertices are in $H$.  These paths can be counted be examining the connections between $v_1$, $v_2$, and all vertices in $H$.  The second contains one vertex in $H$ and one in $V \setminus H$. These paths can be counted by establishing which vertices in $H$ connect to either $v_1$ or $v_2$, and then using the count in $s_2$ of the length two paths from the vertices in $H$ to $v_2$ or $v_1$ respectively.

The $m_9$ count can be increased by an edge insert in two positions.  The first is between the opposite ends of the cycle.  If either $v_1$ or $v_2$ is in $V \setminus H$ then the edge connections can be determined and the count calculated.  If both $v_1$ and $v_2$ are in $H$ then the count of the two two-edge paths that form the cycle must be determined.  These paths will either pass through a vertex in $H$ or a vertex in $V \setminus H$.  The former can be counted by examining the vertices in $H$, and the latter by a lookup to $s_2$.

The second possible position for an edge insert is on the outer path of a cycle that already has an edge through it.  If either $v_1$ or $v_2$ are in $V \setminus H$ calculate the count as follows, summing with an additional calculation considering the vertices reversed. If the vertex connected to the triangle is in $V \setminus H$ then the count can be determined by examining neighbors and their edge connections.  If the vertex not connected to the triangle is in $V \setminus H$ then examine the neighbors.  For those neighbors that are in $V \setminus H$ the count can be determined by examining additional edge connections of neighbors.  For the neighbors in $H$ a lookup $s_2$ is required to completely determine the counts.  If both $v_1$ and $v_2$ are in $H$ then the count is calculated as follows. If all vertices of the structure are in $H$, determine the count by examining edge connections.  If both remaining vertices are in $V \setminus H$ the count can be determined by lookup to $s_5$.  Otherwise, one of the two remaining vertices is in $H$.  This will leave a structure that can be completed and provide a count by using a lookup to $s_2$, or $s_7$

The $m_{10}$ count update is separated by the membership of $v_1$ and $v_2$.  If either vertex is contained in $V \setminus H$, consider $v_1$, then it is possible to determine which vertices connect to $v_1$ and which of these share edges with $v_2$ and each other.  This count can be calculated and the total count can be updated.  If both $v_1$ and $v_2$ are in $H$ then we will sum the values determined in the following three subcases.  First, all four vertices are in $H$.  This count can be determined by examining the edge connections of the vertices in $H$.  If three vertices in $H$ form the correct structure, the count of cliques formed with one vertex in $V \setminus H$ can be determined by a look up to $s_7$.  These counts should be summed for all vertices in $H$ that form the correct structure with $v_1$ and $v_2$.  The final count, with both of the remaining vertices in $V \setminus H$ can be determined by an $s_6$ lookup.

The time complexity for the updates of the stored subgraphs is $O(h^2)$.  Calculations and lookups can be performed in constant time, and subcase calculations can be done independently.  The most complicated subcase count computations involve examination of two sets of connected vertices consecutively that are $O(h)$ in size each.  This results in $O(h^2)$ operations.  The space complexity for our data structure is $O(1)$ for the maintained subgraph counts, $O(m)$ for E, $O(n)$ for the partition to maintain $H$, and $O(mh^2)$ for the dictionaries, because each edge belongs to at most $O(h^2)$ subgraph structures.

\subsection{Subgraph Multiplicity}
The data structure in the previous section only maintains counts of certain subgraph structures.  With the addition of $m$, $n$, and the count of length two paths, where $m$ is the number of edges and $n$ the number of vertices, it is possible to use these counts to determine the counts of all subgraphs on four vertices.  The additional values $m$, $n$, and the count of length two paths can be maintained in constant time per update.  Values for $m$ and $n$ are modified incrementally.  Adding an edge $uv$ will increase the count of length two paths by $d_u + d_v$, the degrees of $u$ and $v$ respectively. Removing the edge will decrease the value by $d_u + d_v - 2$.

\setcounter{MaxMatrixCols}{11}
{\small
\[
\begin{bmatrix}
1&1&1&1&1&1&1&1&1&1&1\\
0&1&2&2&3&3&3&4&4&5&6\\
0&0&1&0&3&3&2&5&4&8&12\\
0&0&0&1&0&0&1&1&2&2&3\\
0&0&0&0&1&0&0&1&0&2&4\\
0&0&0&0&0&1&0&1&0&2&4\\
0&0&0&0&0&0&1&2&4&6&12\\
0&0&0&0&0&0&0&1&0&4&12\\
0&0&0&0&0&0&0&0&1&1&3\\
0&0&0&0&0&0&0&0&0&1&6\\
0&0&0&0&0&0&0&0&0&0&1
\end{bmatrix}
\begin{bmatrix}
q_0\\
q_1\\
q_2\\
q_3\\
q_4\\
q_5\\
q_6\\
q_7\\
q_8\\
q_9\\
q_{10}
\end{bmatrix}
=
\begin{bmatrix}
m_0 = \binom{n}{4}\\
m_1 = m\binom{n-2}{2}\\
m_2 = (n-3)\sum_{v\in V}\binom{\mathrm{degree}(v)}{2}\\
m_3\\
m_4\\
m_5\\
m_6\\
m_7\\
m_8\\
m_9\\
m_{10}
\end{bmatrix}
\]
}

Similar to the matrix for size three subgraphs, we can use the counts of the non-induced subgraphs on the right and the composition of the induced subgraphs to determine the counts of any desired subgraph.

\section{Conclusion}

The work we present here can maintain counts for all 3-vertex directed subgraphs 
$O(h)$ amortized time per update.  This can be done in 
$O(mh)$ space.  For the undirected case, we maintain counts of size-four subgraphs
in $O(h^2)$ amortized time per update and $O(mh^2)$ space.  Although we do not discuss the specifics in this paper, the methodology presented 
can be used to count directed size-four subgraphs with similar complexity. 
These developments open significant possibility for improvement in calculating \emph{graphlet} 
frequencies within Bioinformatics and in ERGM applications for 
social network analysis. 

\bibliographystyle{abbrv}

\end{document}